\documentclass[twocolumn,aps,prb,epsf]{revtex4}
\usepackage{amssymb}
\usepackage{amsmath}
\usepackage{epsfig}
\usepackage{epsf}
\usepackage{array}

\begin{document}

\title{Quasiparticles in high temperature superconductors: consistency of
angle resolved photoemission and optical conductivity}
\author{A J Millis$^{1}$, H. D. Drew$^{2}$}
\affiliation{$^{1}$Department of Physics, Columbia University\\
538 W. 120th St, N.Y., N. Y. 10027\\
$^{2}$ Department of Physics, University of Maryland\\
College Park, MD 20742}

\begin{abstract}
The consistency of angle-resolved photoemission and optical conductivity
experiments on high temperature superconductors is examined. In the limit
(apparently consistent with angle-resolved photoemission data) of an
electron self energy with a weak momentum dependence and a strong frequency
dependence formulae are derived which directly related quantities measured
in the two experiments. Application of the formuale to optimally and
overdoped $Bi_{2}Ca_{2}SrCu_{2}O_{8+\delta}$ shows that the total self
energy inferred from photoemission measurements cannot be interpreted as a
transport scattering rate (in agreement with work of Varma and Abrahams) ,
but that the inelastic part may be so interpreted, if Landau parameter
effects are non-negligible.
\end{abstract}

\maketitle


\narrowtext

\section{Introduction}

The fundamental concept underlying the modern understanding of the physics
of metals is the quasiparticle \cite{Pines61}: the idea that the crucial low
energy eigenstates of interacting electron systems are are sufficiently
similar to conventional electrons that they may be used in standard ways to
calculate transport and other quantities. The utility of the quasiparticle
concept in the case of the high temperature copper oxide superconductors
remains the subject of controversy \cite{review}. Some authors argue that an
intrinsically non-fermi-liquid picture involving unconventional excitations
or a nontrivial critical point is needed. Others assert that a more or less
conventional picture of electrons scattered by some (perhaps somewhat
unconventional) scattering mechanism suffices. Intermediate views exist also
.

Recent improvements in angle-resolved photoemission experiments (for reviews
see \cite{Shen02,Campuzano02}) have provided new insight into this issue.
Detailed measurements of the momentum, temperature and energy dependence of
the electron spectral function \cite%
{Valla99,Valla00,Kaminski01,Valla01,Yusuf02} have demonstrated the
existence, near the fermi surface of optimally doped materials, of
reasonably well defined peaks. \ It seems natural to interpret the peak
position in terms of a quasiparticle dispersion and the peak width in terms
of a quasiparticle scattering rate. The question of the relation between the
scattering rate and dispersion deduced from angle resolved photoemission
experiments and those deduced from the frequency and temperature dependence
of the electrical conductivity immediately arises. A similar question
concerns the relation between the optical and photoemission data and the
predictions of specific models, for example various versions of the 'spin
fermion' model of carriers interacting with spins \cite%
{Jiang96,Stojkovic97,Haslinger01} or the 'marginal fermi liquid' model of
Varma and co-workers \cite{Varma89}. In the existing literature, it is
generally assumed that knowledge of the (low frequency) self energy measured
e.g. via photoemission or calculated via a model determines the frequency
dependent conductivity.

In this paper we examine the issue in more detail.  We show how to compare,
with minimal assumptions, the photoemission and optical data. \ We show that
the scattering rates and mass enhancements deduced from photoemission do not
by themselves describe the high-T$_{c}$ optical data
\cite{Quijada99} in optimally doped
cuprates: first, some portions of the self energy inferred from data must be
discarded (as noted earlier by Varma and Abrahams \cite{Varma01}) and even
after this is done an extra modification (in fermi liquid language a Landau
parameter) must be introduced. \ Implications of these results for our
physical understanding of the cuprates and for attempts at modelling cuprate
properties are outlined.

The rest of this paper is organized as follows. \ Section II summarizes the
formalism needed to dscuss the photoemission data , section III presents the
theory for the conductivity and section IV relates the theory to available
data. Section V is a conclusion.

\section{Photoemission: theory}

The propagation of an electron in a solid is described by the electron Green
function 
\begin{equation}
G(r,t)=<T_{t}\psi (r,t)\psi ^{+}(0,0)>  \label{gdef}
\end{equation}%
where $\psi $ is the electron annihilation operator. \ Photoemission
measurements \cite{Shen02,Campuzano02} allow the determination of the
imaginary part of the Fourier transform of $G$, i.e. the spectral function 
\begin{equation}
A(p,\omega )=ImG(p,\omega )  \label{adef}
\end{equation}%
These measurements are typically interpreted in terms of differences between
the actual $G$ and that corresponding to a reference system in which the
electron propagates without scattering according to some reference
dispersion $\varepsilon _{p}$. \ One defines the self energy $\Sigma $ via 
\begin{equation}
\Sigma (p,\omega )=\omega -\varepsilon _{p}-G^{-1}(p,\omega )  \label{sigdfn}
\end{equation}%
The self energy has both real ($\Sigma ^{^{\prime }}$) and imaginary ($%
\Sigma ^{^{\prime \prime }}$) parts. \ The real part depends on the choice
of reference energy $\varepsilon _{p}$, and the condition $\varepsilon
_{p}+\Sigma (p,\omega =0)=0$ defines the fermi surface. A common choice for
reference dispersion is the dispersion $\varepsilon _{p,band}$ predicted by
the local density approximation to the Kohn-Sham band theory equations; when
band theory results are needed we use the tight binding parametrization of
the LDA band structure determined by Andersen et. al. \cite{Andersen95} and
presented in more detail in the Appendix.

The high $T_{c}$ superconductors have a fundamentally two dimensional
dispersion and a topologically simple fermi surface, so it is convenient to
parametrize momentum by the reference energy $\varepsilon _{p}$ and an
angular coordinate $\theta $ describing position on a surface of constant
reference energy. The observed photoemission spectra\ involve mainly
energies $\omega \lesssim 0.2eV$ in which range the calculated band
dispersion is (especially in high symmetry directions) linear in momentum
(for the direction perpendicular to the fermi surface). For these energies
the observed spectral functions display a reasonably well defined,
reasonably symmetrical peak of approximately Lorentzian form, if $A$ is
measured as a function of $p$ at constant $\omega ,\theta $, ('MDC') but a
rather broad, asymmetric structure if $A$ is measured as a function of $%
\omega $ at constant $\theta ,\varepsilon _{p}$ (EDC). \ The sharpness of
the observed MDC curves implies that in the $\omega <0.2eV$ region where
angle resolved photoemission data are available, the imaginary part of the
self energy is small compared to the range over which $\varepsilon _{p}$
varies. This will be important in the theory of the optical conductivity to
be discussed below. Further, the data suggest  that in the range $\omega
<\omega _{c}\lesssim 0.2eV$, $\Sigma ^{\prime \prime }$ depends reasonably
strongly on $\omega $ and $\theta $ and reasonably weakly on $\varepsilon
_{p}$. However, as we shall see it is likely that at higher energies, beyond
the measurement range, $\Sigma $ depends more strongly on $\varepsilon _{p}$.

It is convenient to introduce a frequency $\omega _{c}$ separating high and
low energy scales and to write 
\begin{equation}
\Sigma (p,\omega )=\Sigma _{low}(\omega ,\theta )+\Sigma _{high}(\omega ,p)
\label{sigdef}
\end{equation}%
with $Im \Sigma _{low}(\omega )=Im \Sigma (\omega )$ for $%
\left\vert \omega \right\vert <\omega _{c}$, $Im \Sigma _{low}=Im%
\Sigma (\omega _{c)}$ for $\left\vert \omega \right\vert >\omega _{c}$ and 
$Re \Sigma _{low}$ the Kramers-Kronig transform of $Im \Sigma
_{low}.$ For energies well below $\omega _{c}$ and momenta not too far from $%
p_{F}$ we linearize $\Sigma _{high}(\omega ,p)$ in $\omega $ and $p-p_{F}$.
(This procedure is unfortunately clumsy--it leads at intermediate stages in
calculations to nonanalytic behavior at $\omega =\omega _{c}$, which of
course cancels from physical quantities but we shall not need to consider $%
\omega =\omega _{c}$ in this paper).

In the low energy region one may therefore write

\begin{equation}
G(p,\omega)=\frac{1}{(\frac{\omega}{Z\left( \theta\right) }-v_{\theta
}(p-p_{F})-\Sigma_{low}(\omega,\theta))}  \label{G-def}
\end{equation}
with 
\begin{equation}
Z(\theta)=\left( 1-\frac{\partial\Sigma_{high}(\omega,\theta)}{\partial
\omega}\right) ^{-1}|_{\omega=0}  \label{Zdef}
\end{equation}
the quasiparticle weight coming from contributions to $\Sigma$ at $%
\omega>\omega_{c},$ and 
\begin{equation}
v_{\theta}=\frac{\partial\left( \varepsilon_{p}+\Sigma_{high}(\omega
<<\omega_{c},p)\right) }{\partial p}|p_{F}  \label{vdef}
\end{equation}
a (possibly angle-dependent) velocity which may differ from the band
velocity if $\Sigma_{high}$ has significant momentum dependence.

Eq \ref{G-def} implies that if measured as a function of $p,$ $A$ has a peak
centered at a momentum $p_{\omega }$ set by 
\begin{equation}
\varepsilon _{p_{\omega }}+\Sigma _{high}(\omega <<\omega _{c},p_{\omega })=%
\frac{\omega -\Sigma _{low}^{^{\prime }}}{Z(\theta )}  \label{pwdef}
\end{equation}%
If the peak is not too far from the fermi level then use of the low energy
approximations shows that the peak position $\delta p_{\omega }$ is given by 
\begin{equation}
\delta p_{\omega }=\frac{\omega -\Sigma ^{\prime }(\theta ,\omega )}{%
v_{\theta }Z(\theta )}  \label{delp}
\end{equation}%
. \ \ If $\Sigma ^{\prime \prime }$ is not too large then one may linearize $%
\varepsilon _{p}$ near $p_{\omega }$, and if $\omega $ is not too large then 
$p$ is close to $p_{F}$ so that we may linearize everything about the fermi
surface, obtaining $\left( \varepsilon _{p}+\Sigma _{high}(\omega <<\omega
_{c},p)\right) =\frac{\omega -\Sigma _{low}^{^{\prime }}}{Z(\theta )}%
+v_{\theta }(p-p_{\omega }),$ so that (in agreement with data) $A$ measured
as a function of momentum (MDC) has an approximately Lorentzian peak of half
width 
\begin{equation}
W(\omega )=\frac{\Sigma _{low}^{\prime \prime }(\theta ,\omega )}{v_{\theta }%
}  \label{W}
\end{equation}%
Similarly, the slope $v_{\omega }^{\ast }=\partial \omega /\partial
p_{\omega }$ of the dispersion curve $\omega =\varepsilon _{p_{\omega }}$ at 
$p=p_{\omega }$ is given by 
\begin{equation}
v_{\theta }^{\ast }=\frac{v_{\theta }}{(1-\partial \Sigma _{low}^{\prime
}/\delta \omega )}  \label{vstar}
\end{equation}%
The quantities $W$ and $v^{\ast }$ are directly measureable, and are
independent of the choice of reference dispersion and of $\ $\ the behavior
of $\Sigma _{high}$. In the analysis of optical conductivity presented in
the next section we shall require their $\omega \rightarrow 0$ limit.

\section{Optical Conductivity: theory}

The conductivity is given by 
\begin{equation}
\sigma _{xx}(i\Omega _{n},T)=\frac{\chi _{jj}^{GI}(i\Omega _{n},T)}{i\Omega
_{n}}  \label{conddef}
\end{equation}%
where $\chi _{jj}^{GI}$ is the gauge invariant current-current correlation
function, which may be expressed in the usual way in terms of electron Green
functions and a vertex operator, $T$. The photoemission data discussed above
indicate that at least in the $\omega <0.2eV$ energy range, the self energy
has negligible dependence on the \textit{magnitude} of the momentum and is
small compared to the range over which $\varepsilon _{p}$ varies. \ In this
case, the usual arguments of fermi liquid theory \cite{Abrikosov63} may be
applied and \ in particular by integrating over the magnitude of the energy
first one finds ($\omega _{n+}=\omega _{n}+\Omega _{n})$ 
\begin{eqnarray}
\chi _{jj}^{GI}(i\Omega _{n},T) &=&\pi T\sum_{\omega _{n}}\int \frac{%
p_{F}(\theta )d\theta }{\left( 2\pi \right) ^{2}}N_{0}v_{x}(\theta )  \notag
\\
&&\frac{\left( sgn(i\omega _{n})-sgn(i\omega _{n+})\right) T_{x}^{i\Omega
}(\theta ,i\omega _{n})}{i\Omega _{n}-\Sigma (\theta ,i\omega _{n+})+\Sigma
(\theta ,i\omega _{n})}  \label{chijj}
\end{eqnarray}%
Here \ we note that the momentum-independence of the self-energy (in the
energy range of interest) implies that the 'bare' current operator is the
'bare' velocity $v_{\theta }$ defined above in Eq \ref{vdef} ('bare' in
quotes because $v_{\theta }$ does depend on $\Sigma _{high}$ and the choice
of reference dispersion).\ $T$ is the vertex operator, defined by

\begin{subequations}
\begin{eqnarray}
T_{x}^{\Omega }(\theta ,\omega ) &=&v_{x}(\theta )+T\sum_{\omega ^{\prime
}}\int \frac{d\theta ^{\prime }}{2\pi }I^{\Omega}(\theta ,\theta ^{\prime };\omega
,\omega ^{\prime })  \label{Tdef} \\
&&\frac{\left( sgn(i\omega ^{\prime })-sgn(i\omega _{+}^{\prime })\right)
T_{x}^{\Omega }(\theta ^{\prime },i\omega )}{i\Omega _{n}-\Sigma (\theta
,i\omega _{+}^{\prime })+\Sigma (\theta ,i\omega ^{\prime })}  \notag
\end{eqnarray}%
with $I$ the generalization to nonzero frequencies of the usual Landau
interaction function (into which we have absorbed the factors of $v$ and $%
p_{F}$).

The vertex function has two purposes: it converts the 'single-particle'
scattering rate and mass enhancements described by $\Sigma $ to a transport
rate and mass enhancement described by a new function $\Sigma _{tr}$ by
suppressing the contribution from 'forward scattering' and it expresses the
'backflow' arising because in an interacting system motion of one electron
affects the motion of others, so the current is not given accurately by the
single particle velocity. Note that if $\partial \Sigma /\partial p\equiv 0$
at all frequencies and momenta then the current operator is given by the
derivative of the reference dispersion $\varepsilon _{p}$ and the backflow
part, $\Lambda $, of the vertex correction vanishes identically. However, a $%
\Lambda $ which is nonnegligible in the frequency range of interest may
arise from a $\partial \Sigma /\partial p$ which is non-negligible only at
frequencies beyond the frequency range of interest. For an explicit example
see Ref\cite{Ioffe89}.

It is perhaps instructive to restate this conclusion in the language of the
quantum Boltzmann equation. There are two nontrivial terms in this equation:
one gives the interaction induced 'feedback' of the excitation of one
particle-hole pair on the behavior of others (i.e. accounts for backflow);
the other is the collision term representing scattering of a quasiparticle
from one state to another. The collision term involves a probability $%
W(p,p^{\prime })$ for scattering an electron from state $p$ to state $%
p^{\prime }$ and the resulting conductivity depends on the structure of $%
W(p,p^{\prime })$: for example, if the scattering is mostly forward ($%
W(p,p^{\prime })$ appreciable only for $p$ near $p^{\prime }$) then the
scattering will have little effect on the conducitivity. The self energy is
proportional to $\int (dp^{\prime })W(p,p^{\prime })\zeta (p^{\prime })$ $\ $%
with $\zeta $ a factor relating to the probability that state $p^{\prime }$
is available as a final state, that any other excitation needed in the
scattering process can be created, etc. Measurement of the self energy by
itself thus does not contain enough information to reconstruct $%
W(p,p^{\prime });$ and in diagrammatic language this information is
contained in the vertex function.

Eq \ref{Tdef} cannot be analysed without further assumptions. We shall
assume that the self energy has two contributions: one coming from low
energies and essentially observable by present-day angle-resolved
photoemission experiments (this is the 'quasiparticle part' of the electron
Green function) and one coming from high energies, not directly observable
in present-day angle-resolved photoemission experiments but contributing
indirectly to low energy physics via the Landau parameter and the velocity
renormalization. We shall further assume, following \cite{Varma01}, that the
low energy contribution to $\Sigma $ consists of two parts: an inelastic
part with a negligible momentum dependence but a signficant frequency
dependence and one arising from a quasistatic scattering highly peaked in
the forward direction. Thus 
\end{subequations}
\begin{equation}
\Sigma (\theta ,\omega )=isgn(\omega )\Gamma _{forward}(\theta )+\Sigma
_{inel}(\omega )+\Sigma _{high}(p,\omega )  \label{sigansatz}
\end{equation}

We shall now write an expression for the low frequency conductivity which
separates the effects of the 'quasiparticle' and high energy contributions.
Eq \ref{sigansatz} implies that the Landau interaction function consists of
two parts: one from the forward scattering contribution, of the form $2\pi
I_{forward}(\theta ;\Omega )\phi (\theta -\theta ^{\prime })$ (independent
of $\omega ,\omega ^{\prime }$ because the scattering is taken to be
quasistatic, and with peakedness in the forward direction specified by $\phi 
$) and one, coming from high energies, which is independent of $\Omega
,\omega ,\omega ^{\prime }$ \ in the frequency range of interest. We define 
\begin{equation}
B(\theta ,i\Omega )=T\sum_{\omega }\frac{i\pi \left( sgn(i\omega
)-sgn(i\omega +i\Omega )\right) }{i\Omega _{n}-\Sigma (\theta ,i\omega
+i\Omega )+\Sigma (\theta ,i\omega )},
\end{equation}%
and $B^{^{\prime }},T^{\prime }$ by 
\begin{equation}
B^{\prime -1}=B^{-1}-I_{forward}
\end{equation}%
\begin{equation}
T^{\prime }=\left( 1-I_{forward}B\right)
\end{equation}%
The vertex $T^{\prime }$ satisfies the simple integral equation 
\begin{equation}
T_{x}^{\prime }(\theta ,\Omega )=v_{x}(\theta )+\int \frac{d\theta ^{\prime }%
}{2\pi }I_{high}(\theta ,\theta ^{^{\prime }})B^{\prime }(\theta ^{\prime
},\Omega )T_{x}^{\prime }(\theta ^{\prime },\Omega )  \label{Tsimple}
\end{equation}%
while the conductivity \textit{per CuO}$_{2}$ plane becomes 
\begin{equation}
\sigma (i\Omega _{n},T)=\frac{1}{i\Omega }\int \frac{p_{F}(\theta )d\theta }{%
2\pi v_{F}(\theta )}v_{x}(\theta )B^{\prime }(\theta ,\Omega )T_{x}^{\prime
}(\theta ,\Omega )  \label{chisimple}
\end{equation}%
We think of the function $B^{\prime }$ as the function $B$ with the forward
scattering contributions removed.

We now consider the low frequency expansion of $\chi _{jj}$. We expect 
\begin{equation}
B^{\prime }(\theta ,\Omega )=\frac{i\Omega }{\Gamma (\theta ,T)}+\frac{%
\Omega ^{2}\Lambda (T)}{\Gamma (\theta ,T)^{2}}  \label{Blow}
\end{equation}%
For example, if the self energy is momentum-independent then (here $\Sigma
_{\pm }=\Sigma (\theta ,\varepsilon \pm \Omega /2)$ 
\begin{eqnarray}
B_{m-i}^{\prime }(\theta ,\Omega ) &=&\int \frac{d\varepsilon }{\pi }  \notag
\\
&&\frac{f(\varepsilon _{-})-f\left( \varepsilon _{+}\right) }{\Omega -\left(
\Sigma _{+}^{\prime }-\Sigma _{-}^{\prime }\right) -i\left( \Sigma
_{+}^{\prime \prime }+\Sigma _{-}^{\prime \prime }\right) }  \label{BMFL}
\end{eqnarray}%
so that 
\begin{align}
\Gamma _{m-i}^{-1}& =\int \frac{d\varepsilon }{\pi }\frac{-\partial
f(\varepsilon )/\partial \varepsilon }{2\Sigma ^{\prime \prime }(\varepsilon
)} \\
\Lambda _{m-i}& =\Gamma _{m-i}^{2}\int \frac{d\varepsilon }{\pi }\frac{%
-\left( 1-\partial \Sigma (\varepsilon )/\partial \varepsilon \right)
\partial f(\varepsilon )/\partial \varepsilon }{\left( 2\Sigma ^{\prime
\prime }(\varepsilon )\right) ^{2}}
\end{align}%
Eq \ref{Blow} implies 
\begin{equation}
T^{\prime }(\theta ;\Omega )=v_{x}(\theta )+i\Omega \int \frac{p_{F}(\theta
^{\prime })d\theta ^{\prime }}{2\pi }\frac{I_{high}(\theta ,\theta
^{^{\prime }})v_{x}(\theta ^{\prime })}{\Gamma _{tr}(\theta ^{\prime })}
\label{Tlow}
\end{equation}

Combining Eqs \ref{Blow}, and \ref{Tlow} yields a low frequency expansion
for the conductivity of the form%
\begin{equation}
\sigma (\Omega )=\sigma _{qp}(\Omega )+\sigma _{LP}(\Omega )  \label{siglow1}
\end{equation}%
with 
\begin{align}
\sigma _{qp}(\Omega )& =\frac{2e^{2}}{\hbar }\int \frac{p_{F}(\theta
)d\theta }{\left( 2\pi \right) ^{2}v_{F}(\theta )}\left[ \frac{%
v_{x}^{2}(\theta )}{\Gamma (\theta )}+i\Omega \frac{\Lambda
(T)v_{x}^{2}(\theta )}{\Gamma (\theta ,T)^{2}}\right]   \label{sigqplow} \\
\sigma _{LP}(\Omega )& =\frac{2e^{2}}{\hbar }i\Omega \int \frac{p_{F}(\theta
)d\theta }{\left( 2\pi \right) ^{2}v_{F}(\theta )}v_{x}(\theta )
\label{sqlplow} \\
& \int \frac{p_{F}(\theta ^{\prime })d\theta ^{\prime }}{\left( 2\pi \right)
^{2}}\frac{I_{high}(\theta ,\theta ^{^{\prime }})v_{x}(\theta ^{\prime })}{%
\Gamma (\theta ^{\prime })}  \notag
\end{align}%
where $\sigma _{qp}$ is the contribution obtained from the quasiparticle
scattering and dispersion and $\sigma _{LP}$ arises from the Landau or
backflow renormalization. Observe that the Landau renormalization affects
the first frequency correction to the conductivity, but not the dc value.

It is very convenient to write this expression in terms of an
inverse  'transport'
mean free path $W$ and a transport velocity defined analogously to Eq \ref{W}: 
\begin{equation}
W_{tr}(\theta )=\Gamma (\theta )/v_{F}(\theta )  \label{Wtr}
\end{equation}%
\begin{equation}
v_{tr}^{\ast }(\theta )=v(\theta )/\Lambda (\theta )  \label{v*tr}
\end{equation}%
Then one has 
\begin{align}
\sigma _{qp}(\Omega ,T)& =\frac{2e^{2}}{\hbar }\int \frac{p_{F}(\theta
)d\theta }{\left( 2\pi \right) ^{2}}\left( \frac{v_{x}(\theta )}{v(\theta )}%
\right) ^{2}  \notag \\
& \left( \frac{1}{W_{tr}(\theta )}+\frac{i\Omega }{v^{\ast }(\theta
)W_{tr}(\theta )^{2}}\right)   \label{sigqp2} \\
\sigma _{LP}& =i\Omega \int \int \frac{p_{F}(\theta )d\theta }{\left( 2\pi
\right) ^{2}}\frac{p_{F}(\theta ^{\prime })d\theta ^{\prime }}{\left( 2\pi
\right) ^{2}}  \notag \\
& \frac{\frac{v_{x}(\theta )}{v_{F}(\theta )}I_{high}(\theta ,\theta
^{^{\prime }})\frac{v_{x}(\theta ^{\prime })}{v_{F}(\theta )}}{W_{tr}(\theta
)W_{tr}(\theta ^{\prime })}  \label{sigLP2}
\end{align}

In particular, the dc limit of the conductivity is 
\begin{equation}
\sigma _{dc}(T)=\frac{2e^{2}}{\hbar }\int \frac{p_{F}(\theta )d\theta }{%
\left( 2\pi \right) ^{2}}\left( \frac{v_{x}(\theta )}{v(\theta )}\right) ^{2}%
\frac{1}{W_{tr}(\theta )}  \label{sigdc}
\end{equation}%
and is given entirely in terms of fermi surface geometry and the transport
mean free path.

In the absence of Landau renormalization, the imaginary part $\sigma
^{\prime \prime }\rightarrow \sigma _{qp}^{\prime \prime }$ given by 
\begin{equation}
\lim_{\Omega \rightarrow 0}\sigma _{qp}^{\prime \prime }(\Omega )=\frac{%
2e^{2}}{\hbar }i\Omega \int \frac{p_{F}(\theta )d\theta }{\left( 2\pi
\right) ^{2}}\left( \frac{v_{x}(\theta )}{v(\theta )}\right) ^{2}\frac{%
i\Omega }{v^{\ast }(\theta )W_{tr}(\theta )^{2}}  \label{sigqp}
\end{equation}

In experimental analyses of optical conductivity it is conventional (see,
e.g. \cite{Puchkov96}) to define an optical mass and scattering rate via 
\begin{align}
\Gamma _{opt}(\Omega )& =KRe \sigma (\Omega )^{-1}  \label{optrate} \\
\frac{m^{\ast }}{m}_{opt}(\Omega )& =-\frac{K Im\sigma (\Omega )^{-1}}{%
\Omega }  \label{optmass}
\end{align}%
where $K$ is a constant related to the optical spectral weight in the
frequency range of interest. The values of $\Gamma $ and $m^{\ast }/m$
depend on the value used for $K$ , leading to ambiguity in the values of $%
\Gamma _{opt}$ and $m^{\ast }/m_{opt}$ similar to the ambiguity in the
single-particle self energy arising from uncertainty as to the correct
choice of reference velocity. One quantity which is independent of the
choice of $K$ is the ratio 
\begin{equation}
\lim_{\Omega \rightarrow 0}\Gamma _{opt}^{\ast }=
\frac{i\Omega Re \sigma }{ Im \sigma }  \label{gamstar}
\end{equation}

The formulae given above involve quantities defined in the $\Omega
\rightarrow 0$ limit. One may consider extending the analysis to higher
frequencies, but our lack of knowledge of the vertex function renders such
an analysis problematic.

Eqs \ref{sigqp2}, \ref{sigLP2} are our principal results. They show that
measurement of the transport mean free path and quasiparticle velocity
predict the dc conductivity and its first frquency derivative only if Landau
parameter effects are negligible. Thus if measurements of $W$ and $v^{\ast }$
are available, comparison of Eq \ref{sigqp2} to data will show whether
Landau parameter effects are important.

\section{Optical and Photoemission Data in High T$_{c}$ superconductors:
analysis and consistency with quasiparticle picture}

In this section we use the formulae derived above to investigate the
relation between the photoemission and optical spectra in  optimally
and over-doped $%
Bi_{2}Sr_{2}CaCu_{2}O_{8+\delta }$, and in particular to determine whether
the scattering rates inferred from photoemission experiments may be
interpreted as transport rates, and to estimate the value of any Landau
renormalization. We have selected this material because extensive
photoemission and optical data are available. In other high-$T_{c}$
materials insufficient information exists to perform the analysis at present.

Experiments show that in optimally doped BSCCO the fermi surface is to
reasonable approximation a circle of radius $p_{F}=0.71$\AA $^{-1}$ centered
at the $(\pi ,\pi )$ point. The quasiparticle velocity $v^{\ast }$=1.8eV-%
\AA\ with negligible variation around the fermi surface (in the normal
state), and the 'MDC full width' $2W(\theta ,T,\varepsilon )$ is reasonably
well represented by \cite{Valla99,Valla00} 
\begin{equation}
2W(\theta ,T,\varepsilon )=\Gamma _{0}\max (\varepsilon ,\pi T)+\Gamma
_{1}(1+\cos (4\theta ))+\Gamma _{2}  \label{Wopt}
\end{equation}%
with $\Gamma _{0}=8\times 10^{-5}\left[ \text{\AA }^{-1}K^{-1}\right] ,$ $%
\Gamma _{1}=0.05$\AA $^{-1}$ $\Gamma _{2}=0.01$\AA $^{-1}$ and $\theta =0$
at the antinodal point$(0,\pi )$. Comprehensive data from
other groups are not available as of this writing but we
note that the zone-diagonal $\omega=0$ MDC widths reported in 
Ref \cite{Kaminski01} are very close to the zone diagonal numbers
obtained from the formula above. 

Let us first make the assumption that
vertex corrections are negligible. Then from Eqs \ref{sigdc} and \ref{sigqp}
we obtain ($c$ is the mean interplane distance) 
\begin{align}
\sigma _{dc}& =\frac{e^{2}}{\hbar c}\frac{p_{F}}{2\pi }\int d\varepsilon 
\frac{\partial f}{\partial \varepsilon }I_{1}(\varepsilon ,T)
\label{sigdcopt} \\
\lim_{\Omega \rightarrow 0}\frac{\sigma _{qp}^{\prime \prime }(\Omega )}{%
\Omega }& =\frac{e^{2}}{\hbar c}\frac{p_{F}}{2\pi }\frac{1}{v^{\ast }}\int
d\varepsilon \frac{\partial f}{\partial \varepsilon }I_{2}(\varepsilon ,T)
\label{sig2opt}
\end{align}

\bigskip The two integrals are 
\begin{align}
I_{1}& =\int \frac{d\theta }{2\pi }\frac{1}{W(\theta ,T,\varepsilon )}= 
\notag \\
& \frac{1}{\sqrt{\left( \Gamma _{0}\max (\varepsilon ,\pi T)+\Gamma
_{2}\right) ^{2}+2\left( \Gamma _{0}\max (\varepsilon ,\pi T)+\Gamma
_{2}\right) \Gamma _{1}}}  \label{I1} \\
I_{2}& =\int \frac{d\theta }{2\pi }\frac{1}{W(\theta ,T,\varepsilon )^{2}}= 
\notag \\
& \frac{\left( \Gamma _{0}\max (\varepsilon ,\pi T)+\Gamma _{2}\right)
+\Gamma _{1}}{\left( \Gamma _{0}\max (\varepsilon ,\pi T)+\Gamma
_{2}+2\Gamma _{1}\right) ^{3/2}\left( \Gamma _{0}\max (\varepsilon ,\pi
T)+\Gamma _{2}\right) ^{3/2}}  \label{I2}
\end{align}

Let us suppose first ('Case A') that the entire observed photoemission
linewidth may be interpreted as a transport scattering rate. Then from Eq %
\ref{I1} we obtain the resistivities listed in the second row of Table I.
These are plainly higher than the measured resistivities, listed in the
third row of Table I. As an alternative assumption we may follow Varma and
Abrahams \cite{Varma01} and argue that only the inelastic ($\Gamma _{0}$)
should be interpreted as a contribution to the transport rate. In this case
('Case B') we obtain the numbers shown in the second row of Table I, which
as previously noted by Varma and Abrahams \cite{Varma01}
are in reasonably good agreement with data \cite{Quijada99}

\begin{table}[tbp] \centering%
\begin{tabular}{|c|c|c|c|}
\hline
& 100K & 200K & 300K \\ \hline
$\rho$: Case A & $162$ & 232 & 300 \\ 
$\rho:$ Case B & $60$ & 120 & 180 \\ 
$\rho:$ Data & $75$ & 130 & 240 \\ \hline
\end{tabular}
\caption{$\rho$[$\mu\Omega$ cm] calculated for optimally doped
$Bi_2Ca_2SrCu_2O_{8+\delta}$ at temperatures indicated,
using photoemission data  \cite{Valla00}  assuming (A) directly measured MDC width
(B) Only $T$
and $\omega$-linear parts, and compared to data \cite{Quijada99}.
\label{rhoopt}}%
\end{table}%

We now proceed further with the analysis and consider the leading correction
to the imaginary part of the conductivity. These data are conveniently
presented in terms of the 'optical scattering rate $\Gamma _{opt}^{\ast }$.
\ Numerical evaluation of Eq \ref{I2} assuming a velocity which is
temperature and frequency independent leads to the results shown in Table
II. Photoemission information on the temperature dependence of the velocity
is lacking, but available data \cite{Valla01} do not suggest a strong
frequency dependence in the low frequency regimes relevant to this
calculation We note that the results are somewhat sensitive to the precise
assumptions made. To demonstrate this point we consider two sub cases of
'Case B: above--where we retain only the inelastic portion of the scattering
rate and (Case B'") where we retain also the $\Gamma _{2}=0.01A^{-1}$
angle-independent offset. Table II summarizes our calculated results.

\begin{table}[tbp] \centering%
\begin{tabular}{|l|l|l|l|}
\hline
& \textbf{100K} & \textbf{200K} & \textbf{300K} \\ \hline
$\Gamma ^{\ast }:Case$ A & 100 & 160 & 210 \\ 
$\Gamma ^{\ast }:$ $Case$ B & 46 & 92 & 140 \\ 
$\Gamma ^{\ast }:$ $Case$ B' & 64 & 110 & 155 \\ \hline
$\Gamma ^{\ast }:$ $data$ & 20 & - & 80 \\ \hline
\end{tabular}
\caption{Effective scattering rate $\Gamma^*$ [meV] calculated  for optimally doped $Bi_2Sr_2CaCu_2O_{8+\delta}$
from photoemission data  \cite{Valla00}  assuming (A) the entire 
photoemission scattering rate contributes to the
conductivity (B) only the inelastic part contributes and (B') that 
both the inelastic
and the offset at the zone diagonal $\Gamma_2$ parts contribute, 
and compared to
data \cite{Quijada99}
\label{gamstaropt}}%
\end{table}%

$\ $

The dc resistivities in 'Case A' are much too large to be relevant to
optimally doped BSCCO. We conclude that at least the strongly
angle-dependent contribution to $W$ cannot correspond to a transport rate.
This conclusion was previously stated by Varma and Abrahams \cite{Varma01},
who attributed the $\Gamma_{1}$ term to impurities situated far from the $%
CuO_{2}$ planes. The ubiquity of the large MDC widths in this region of the
fermi surface suggests to us that this explanation is untenable; however the
conclusion that the width does not correspond to a transport rate seems
inescapable. Interestingly, the dc conductivities in Case B seem to agree
reasonably with the observed resistivity, suggesting that the 'inelastic'
part of the angle resolved photoemission MDC width does correspond to a
transport rate. However, using these parameters in the quasiparticle
formulae strongly overestimates the renormalized optical scattering rate,
especially at lower $T$, suggesting that there is a significant Landau
renormalization of the conductivity.

For other doping levels the comparison is more difficult to undertake at
this stage, because the photoemission data are less extensive. A recent
paper \cite{Yusuf02} presents evidence that in an overdoped sample of $%
Bi_{2}Sr_{2}CaCu_{2}O_{8+\delta}$, $W(\theta,\omega=0)$ is only weakly
dependent on angle, but \ varies more rapidly than linearly with
temperature, being about $W=0.02$\AA $^{-1}$ at 70K rising to $0.04$\AA $%
^{-1}$\ at $120K$ and $.057$\AA $^{-1}$ at $160K$.\ \ The fermi surface
radius (measured from the $(\pi,\pi)$ point) is slightly larger ($0.78$\AA $%
^{-1}$ vs $0.72$\AA $^{-1}$ in the optimally doped material studied in \cite%
{Valla00}. The frequency dependence of $W$ is still found to be linear (at
least at very low $T$). Converting from the units of \cite{Yusuf02} to the
conventions of this paper yields, in convenient units%
\begin{equation}
W_{OD}(\omega,T\rightarrow0)\left[ \mathring{A}^{-1}\right] =1.2\times
10^{-5}\omega\lbrack K]  \label{Wod}
\end{equation}
\qquad

The crossover between $\omega$-dominated and $T-$dominated regimes is not
discussed, however as can be seen from Eq \ref{Wod}, for a frequency
corresponding to $600K$ the frequency dependent contribution is only $\Delta
W_{OD}=7.2\times10^{-3}$\AA $^{-1}$ negligible compared to the dc scattering
rate. Therefore we may to reasonable accuracy simply neglect the $\omega $%
-dependence of the scattering rate, and use a Drude model. Our results are
given in Tables III and IV.

\begin{table}[tbp] \centering%
\begin{tabular}{|l|l|l|l|}
\hline
& 70K & 120K & 160K \\ \hline
$\rho-A$ & 43 & 85 & 121 \\ 
$\rho-data$ & 58 & 70 & 88 \\ \hline
\end{tabular}
\caption{$\rho$[$\mu\Omega$ cm] calculated for overdoped
$Bi_2Ca_2SrCu_2O_{8+\delta}$ at temperatures indicated,
using directly measured MDC width  \cite{Yusuf02}  
and compared to resistivity data from the same paper.}

\end{table}%

\begin{table}[tbp] \centering%
\begin{tabular}{|l|l|l|l|}
\hline
& 70K & \textbf{120K} & \textbf{160K} \\ \hline
$\Gamma ^{\ast }-A$ & 36 & 72 & 102 \\ 
$\Gamma ^{\ast }-data$ & 19 & 28 & 35 \\ \hline
\end{tabular}
\caption{Effective scattering rate $\Gamma^*$ [meV] 
calculated  for  overdoped $Bi_2Sr_2CaCu_2O_{8+\delta}$
from measured MDC widths  \cite{Yusuf02}   and compared to
data \cite{Santanders03}}
\end{table}%
$\ \ \ $\newline

Here, as noted by the authors of Ref \cite{Yusuf02} the photoemission and
resistivity data appear to have an inconsistent temperature dependence.
Also, the optical scattering rate is again underpredicted, suggesting the
importance of a Landau parameter.

For underdoped materials sufficient photoemission data does not yet exist to
make the comparison feasible. Determining the behavior of the Landau
parameter with doping would be very important.

\section{Conclusion}

We have presented a precise and reasonably model-independent method for
comparing the photoemission and optical scattering rates and mass
enhancements, and have applied the method to optimally doped and overdoped $%
Bi_{2}Ca_{2}SrCu_{2}O_{8+\delta }$. \ Our method provides relations between
directly measured quantities and therefore provides an unambiguous test of
whether the 'MDC width' measured in angular resolved photoemission
experiments corresponds to a transport mean free path. In agreement with
previous authors, we find that it does not. The discrepancy is particularly
severe in the case of optimally doped $Bi_{2}Ca_{2}SrCu_{2}O_{8+\delta }$
where use of the full MDC width grossly overpredicts the resistivity. We
conclude, in agreement with previous authors \cite{Varma01} that the
broadening of the photomission spectra in the vicinity of the $(0,\pi )$
point of the fermi surface should not be regarded as a contribution to the
transport part of the self energy. The authors of Ref \cite{Varma01} argued
that the large broadening in this part of the zone arises from elastic
scattering by out of plane impurities. In our view the ubiquity of the
zone-corner broadening in cuprate materials argues instead in favor of an
intrinsic, probably many-body origin to the phenomenon;
understanding why it does not remains a very challenging
theoretical problem. However, assuming
that this broadening enters transport in the usual way is inconsistent with
data. 
In our view the apparent  irrelevance of the large zone-corner self-energy
to the low frequency transport
casts doubt on the attempts to describe transport and
optical propeties with a 'spin-fermion' model \cite{Stojkovic97,Haslinger01}, 
because in these models it is precisely the zone-corner scattering rate
which is taken to be crucial for the conductivity. HOwever, it
is possible to find parameter regimes in spin-fermion models
for which the scattering is not so strongly angle-dependent and
reasonable (modulo Landau-parameter effects) fits to the conductivity
may be achieved \cite{Haslinger01}.

Our main new finding is that even if one is selective in the part of the
photoemission data one interprets as giving rise to a transport rate (for
example by selecting only the 'inelastic' part, agreement between
calculation and experiment cannot be obtained unless a 'Landau parameter'
(corresponding to an interaction-induced vertex correction to the
conductivity) is introduced. The importance of this vertex correction casts
doubt on most of the existing calculations of the frequency and temperature
dependent conductivity, which neglect vertex corrections. This conclusion
may be stated in a different way. If (as, for example, was very elegantly
done in Ref \cite{Quijada99}) a model self energy is constructed which
reproduces (\textit{without vertex corrections}) the conductivity spectrum,
this self energy will necessarily fail to fit the photoemission spectrum. \
Determining the doping dependence of the vertex correction factor is an
important topic for future research. We suspect that this must be large,
because the low frequency optical spectral weight (which is closely related
to the $\Gamma ^{\ast }$ discussed above, displays a strong doping
dependence whereas the observed low energy photoemission velocity does not.
We also note that information on the temperature and frequency dependence of
the photoemission-determined quaisparticle velocity would considerably help
in making this comparison precise.

\textit{Acknowledgements:} H. D. D. was supported
in part by NSF-DMR-0070959.  A. J. M was supported in part by NSF DMR
00081075, in part by the Institute for Theoretical Physics in part by the
Department of Energy at Brookhaven National Laboratory and in part by the
ESPCI (France), and thanks the ESPCI for hospitality. We thank Eilhu
Abrahams, Nicole Bontemps, J. C. Campuzano, L. B. Ioffe and Andres Santander
for helpful discussions and Nicole Bontemps and A. Santanders for provision
of unpublished data.

\medskip 

\centerline{\large{{\bf  Appendix: Band Theory}}}

\medskip 

The natural choice for the reference dispersion is that given by band
theory. There is general agreement that the dispersion in a single $CuO_{2}$
plane of a high $T_{c}$ superconductor is given to a good approximation by 
\begin{eqnarray}
\varepsilon _{LDA}(p) &=&-2t(\cos (p_{x}a)+\cos (p_{y}a))  \notag \\
&&+4t^{\prime }\cos (p_{x}a)\cos (p_{y}a) \\
&&-2t^{\prime \prime }(\cos (2p_{x}a)+\cos (2p_{y}a))  \notag
\end{eqnarray}%
The best choice of the parameters is a subtle issue \cite{Andersen95}.
Especially, the behavior in the vicinity of the zone corners $(0,\pi ),(\pi
,0)$ depends sensitively on details, but the zone-diagonal velocity is
reasonably robust, varying between about $3.8-4.1eV-A$ depending on
calculation and precise doping. We adopt here $t=0.38eV$, $t^{\prime }=0.32t$
and $t^{\prime \prime }=0.5t$ \cite{Andersen95} implying a zone-diagonal
velocity $\partial \varepsilon _{LDA,p}/\partial p=3.9-4.1\;eV-A$ with the
variation arising mainly from nonlinearities in the dispersion. \ The fermi
line parameter $p_{FS}\approx 0.7A^{-1}$. It is also sometime convenient to
define the kinetic energy $K$ via 
\begin{equation}
K=2\int \frac{d^{2}p}{(2\pi )^{2}}v_{p,x}^{2}\delta (\varepsilon _{p}-\mu
)=vp_{FS}/2\pi 
\end{equation}%
The band 'kinetic energy $K$ is about $0.35eV$ corresponding to an average $v
$ of about $3eV-A$.

\newpage

\end{document}